\documentclass[aps,pre,reprint ,a4paper,superscriptaddress,groupedaddress]{revtex4-1}  
\pdfoutput=1

\usepackage[T1]{fontenc}
\usepackage[latin1]{inputenc}
\usepackage[english]{babel}

\usepackage{amsmath, amssymb, amsthm}  
\usepackage{bm, bbm}

\usepackage[]{graphicx}

\usepackage[colorlinks = true, linkcolor = blue, urlcolor  = blue, citecolor = blue,       anchorcolor = blue]{hyperref} 

\usepackage[amssymb]{SIunits}  
\usepackage{enumerate} 
\usepackage{dcolumn}   
\usepackage{lineno}

\usepackage{comment}
\usepackage{slashed}    
\usepackage{mathtools}  

\usepackage{tikz}
\usetikzlibrary{matrix}
\usetikzlibrary{positioning} 
\usetikzlibrary{external}

\newcommand{\uptau}{\tau}
\newcommand{\acos}{\arccos}
\DeclareMathOperator{\diag}{diag}
\DeclareMathOperator{\sgn}{sgn}
\let\Re\relax\DeclareMathOperator{\Re}{Re}

\newcommand{\symcen}[2]{\ooalign{$\phantom{#1}$\cr\hfil$#2$\hfil\cr}}

\newcommand{\iu}{\mathfrak{i}}
\newcommand{\ex}{ {e}}
\newcommand{\du}{ {d}}
                      
\newcommand{\mat}[1]{\bm #1}
\renewcommand{\vec}[1]{\bm #1}
\newcommand{\abs}[1]{\lvert #1 \lvert}
                         
\newcommand{\tfra}[1]{\tfrac{1}{#1}}

\addto\captionsenglish{}
\addto\captionsenglish{}

\newcommand{\dcoef}{D}
\newcommand{\ori}{\phi}
\newcommand{\ter}{\psi}
\newcommand{\sbTheta}{\mathrlap{\bm\Theta}{\slashed{\phantom O}}}

\begin{document}

\title{Different approach to the modeling of nonfree particle diffusion}
\author{Niels Buhl}
\email{nbuhl@phys.au.dk}
\affiliation{School of Physics and Astronomy, University of Nottingham, Nottingham NG7 2RD, United Kingdom}


\begin{abstract}
A new approach to the modeling of nonfree particle diffusion is presented. The approach uses a general setup based on geometric graphs (networks of curves), which means that particle diffusion in anything from arrays of barriers and pore networks to general geometric domains can be considered and that the (free random walk) central limit theorem can be generalized to cover also the nonfree case. The latter gives rise to a continuum-limit description of the diffusive motion where the effect of partially absorbing barriers is accounted for in a natural and non-Markovian way that, in contrast to the traditional approach, quantifies the absorptivity of a barrier in terms of a dimensionless parameter in the range 0 to 1. The generalized theorem gives two general analytic expressions for the continuum-limit propagator: an infinite sum of Gaussians and an infinite sum of plane waves. These expressions entail the known method-of-images and Laplace eigenfunction expansions as special cases and show how the presence of partially absorbing barriers can lead to phenomena such as line splitting and band gap formation in the plane wave wave-number spectrum.          
\end{abstract}

\date{\today}
\maketitle

Particle diffusion in the presence of barriers that each exhibit a combination of reflective, transmittive, and absorptive properties occurs ubiquitously in nature 
(see, e.g., Refs.~\cite{Latour1994,Hurlimann1994,Wang2006}) and the ability to properly model this phenomenon is as such of great practical and theoretical importance. Starting from the fundamental idea (due to Ref.~\cite{Einstein1905}) of dividing the observation period into equal-duration time intervals of such a magnitude that the corresponding displacements of a particle can be considered statistically independent, one gets a random walk description of the diffusive motion. If no barriers are present, one can apply the (free random walk) central limit theorem to this and immediately get the fundamental quantity, the continuum-limit propagator, with which the diffusive motion can be described analytically using only one free parameter (the diffusion coefficient). 
To get a similar continuum-limit description of the diffusive motion  when barriers are present, it has so far been necessary to derive (from the random walk description) and solve an initial-boundary value problem. 
With this traditional approach one has to account for the effect of a partially absorbing barrier in a Markovian way using a Robin boundary condition and quantify the absorptivity of a barrier in terms of a nondimensionless Robin boundary condition coefficient in the range $0$ (nonabsorbing) to infinity (completely absorbing)  (see, e.g., Ref.~\cite{Redner2001}).

In this Rapid Communication we introduce a different approach to the modeling of nonfree particle diffusion by showing that the (free random walk) central limit theorem can be generalized to cover also the nonfree case. This makes it possible to get a continuum-limit description of the diffusive motion where the effect of partially absorbing barriers is accounted for in a natural and non-Markovian way that lies beyond what one can do with the initial-boundary value problem framework. 
The approach proceeds by first reducing the problem to the modeling of diffusive motion of points, called particle points, on an appropriately chosen geometric graph. 
By a geometric graph we mean a geometric object composed of noncoinciding points, called vertices, and equal-length curves, called edges, such that each end point of each edge coincides with a vertex, each vertex coincides with at least one and at most finitely many edge end points, and no edge has coinciding end points. 
If, e.g., the problem concerns particle diffusion perpendicular to one or more parallel plane barriers, then the appropriate geometric graph is composed as a chain of line segments and the particle points are the particles' projections hereon. 
For particle diffusion over medium or long distances in a confined region with a networklike topology, such as the air space in a mammalian lung or the void space in a porous material, the appropriate geometric graph could, e.g., represent the central or medial axis through the airways or the void space, respectively, and the particle points the particles' projections hereon. 
For particle diffusion in a general confined region along a direction that cannot be treated independently of the two orthogonal directions, the appropriate geometric graph is a structured grid approximating the confined space and the particle points represent the particles' projections hereon if the grid is planar and the particles themselves if the grid is three dimensional. 
The observation period is then divided into equal-duration time intervals whose magnitude, called the step duration, is such that the corresponding displacements of a particle point can be considered to be statistically independent and to have a standard deviation, called the step length, that is equal to a nonunity unit fraction of the edge length.  
The random walk process giving the random walk description of the diffusive motion is then obtained by successively adding random displacements head to tail on the geometric graph (one each time a step duration has elapsed) such that the distances from the tail of the first random displacement to the end points of the edge on which it is placed are multiples of the step length and each random displacement is unbiased and one step length long.  
The addition of random displacements is continued until a vertex is reached. The addition is then terminated with a probability that is equal to the absorption probability associated with this vertex. If termination does not occur, then the addition is continued until a neighboring vertex is reached. The addition is then terminated with a probability that is equal to the absorption probability associated with this vertex. If termination does not occur, then the addition is continued until a neighboring vertex is reached and so on. 
Note that the random walk process is non-Markovian if there is at least one vertex whose associated absorption probability is strictly between $0$ and $1$ since the probability of continuing after reaching such a vertex will depend on how the vertex is reached (if the addition continues after reaching such a vertex, then it will continue with probability $1$ if the vertex is reached again before a neighboring vertex is reached). 
The discrete propagator for the random walk process is  obtained in analytic form by noting that it is equal to a sum whose terms can be expressed in terms of the discrete propagator for a free random walk process. This immediately leads to two general analytic expressions for the continuum-limit propagator (an infinite sum of Gaussians and an infinite sum of plane waves) in which the diffusion coefficient (the  step length squared divided by twice the step duration) and the complements of the absorption probabilities are the only free parameters.

To begin, let $\mathcal G$ be the geometric graph appropriate for the given problem. Let   $(v_\ell)_{\ell=1}^V$ be the vertices, $(e_i)_{i=1}^E$ be the edges, and $L$ be the edge length of $\mathcal G$. Let $d_\ell$ be the degree of $v_\ell$, i.e., $d_\ell$ is the number of edges that have an end point that coincides with $v_\ell$. 
A direction is given to each edge. Let $\ori_i$ and $\ter_i$ be the two positive integers such that $v_{\ori_i}$ coincides with that end point of $e_i$ that $e_i$'s direction is away from and $v_{\ter_i}$ coincides with that end point of $e_i$ that $e_i$'s direction is towards.
Let $\uptau$ be the step duration and $\sigma$ be the step length (such that $L/\sigma$ is an integer strictly greater than $1$). Let $1-c_\ell$ be the absorption probability (and $c_\ell$ be the continuation probability) associated with $v_\ell$.

Let $K(y,x,t)_{ji}$, where  $i,j={1,\dotsc,E}$, $t=0,\uptau,2\uptau,\dotsc$, and  $x,y=\sigma,2\sigma,\dotsc,L-\sigma$ be the discrete propagator for the random walk process, i.e., $K(y,x,t)_{ji}$ is the sum of the probabilities of all  trajectories (i.e., realizations of the process) of duration $t$ that goes from $(i,x)$ to $(j,y)$. 
In other words, $K(y,x,t)_{ji}$  describes the probability that a particle point that starts at $(i,x)$ is at $(j,y)$ after diffusing for a period of duration $t$. The notation $(i,x)$ refers to  the position on $e_i$ whose distance from $v_{\ori_i}$ is equal to $x$. 
If $(\ell_1,i_1,\ell_2,i_2,\dotsc,\ell_k)$ is a $k$-vertex path for which $\ell_1=\ter_i$ and  $\ell_k=\ori_{j},\ter_j$, then we consider the sum of the probabilities of all  trajectories of duration $t$ that goes from $(i,x)$ to $v_{\ell_1}$ without visiting any vertices on the way, makes a non-negative number of excursions, goes via $e_{i_1}$ to $v_{\ell_2}$, makes a non-negative number of excursions, goes via $e_{i_2}$ to $\ldots$ $v_{\ell_k}$, makes a non-negative number of excursions, and goes to  $(j,|\delta_{\ter_j\ell_k}L-y|)$ without visiting any vertices on the way. 
A segment of a trajectory is called an excursion if it goes from a vertex back to the same vertex without visiting any vertices on the way. 
This sum is equal to         $({2c_{\ell_k}}/{d_{\ell_k}})\dotsm({2c_{\ell_1}}/{d_{\ell_1}})q_k(y,x,t)$, where $2^kq_k(y,x,t)$ is the value that this sum attains if $\mathcal G$ has only one edge ($E=1$) and $c_1,c_2=1$.    
To see this, note that for each such trajectory in the special case $E=1$ and $c_1,c_2=1$ there are    $(d_{\ell_k})^{\varepsilon_k}\dotsm(d_{\ell_1})^{\varepsilon_1}$ such trajectories that each has a probability of     $c_{\ell_k}(1/d_{\ell_k})^{\varepsilon_k+1}\dotsm c_{\ell_1}(1/d_{\ell_1})^{\varepsilon_1+1}w$ in the general case, where $w$ is the probability of (and $\varepsilon_1,\dotsc,\varepsilon_k$ are the relevant numbers of excursions made by) the considered special case trajectory.
The sum of the probabilities of all trajectories of duration $t$ that goes from $(i,x)$ to $(j,y)$ can thus be written [$q_0(y,x,t)$ is the sum of the probabilities of all trajectories of duration $t$ that goes from $(i,x)$ to $(i,y)$ without visiting any vertices] 
\begin{align}
K(y,x,t)_{ji}=\delta_{ji}\, q_0(&y,x,t) +\sum_{k>0}\sum_{(\ell_1,\dotsc,\ell_k)}\tfrac{2c_{\ell_k}}{d_{\ell_k}}\dotsm\tfrac{2c_{\ell_1}}{d_{\ell_1}}\nonumber\\
\times&\,q_k(|\delta_{\ter_{j}\ell_k}L-y|,|\delta_{\ori_i \ell_1}L-x|,t),\label{eq:firstqnexpression}
\end{align}
where $\delta_{ij}$ is the Kronecker delta function and the inner sum is over all $k$-vertex paths $(\ell_1,\dotsc,\ell_k)$ for which  $\ell_1=\ori_i,\ter_i$ and  $\ell_k=\ori_j,\ter_j$.  
A $k$-vertex path is a  sequence of vertex and edge indices  $({\ell_1},i_1,{\ell_2},i_2,\dotsc,{\ell_k})$ such that     $\{\ori_{i_{m}},\ter_{i_m}\}=\{\ell_{m},\ell_{m+1}\}$ for each $0<m<k$.   	
\begin{figure}[!htp]
\vspace{0.25cm}
\centering
\includegraphics[]{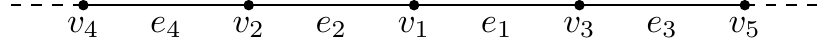}
\caption{\label{fig:line}An infinite chain (i.e., an infinite geometric graph where  each vertex has degree $2$).
}
\end{figure}

To obtain an analytic expression for the function  $q_k(y,x,t)$ we assign, exclusively in this paragraph, $\mathcal G$ to be an infinite chain and   $(v_\ell)_{\ell=1}^\infty$,   $(e_i)_{i=1}^\infty$, the edge directions, and the continuation probabilities to be such that $\ori_i=i$,     $\ter_i=i-(-1)^{i}2+\delta_{i2}$ (see Fig.~\ref{fig:line}), and $c_1,c_2,\dotsc=1$. Let then      $g_n(y,x,t)=K(y,x,t)_{(|2n|+\theta(n))1}$, where $n$ is an integer and  $\theta(\cdot)$ is the Heaviside step function.     
Let furthermore   $q_{k}(y,x,t)=q_{-k}(L-y,L-x,t)$ if $k<0$. It then follows from \eqref{eq:firstqnexpression} that  
\begin{align*} 
g_{n}(y,x,t)=\sum_{k}^{}a_{nk}\,q_{k}(y,x,t)+ b_{nk}\,q_{k}(L-y,x,t), 
\end{align*} 
where $a_{n 0}=\delta_{n 0}$ and $a_{n k}=\alpha_{n-\sgn(k),\abs k}$, $k\ne 0$, together with $ b_{n 0}=0$ and $b_{n k}=\alpha_{n,\abs k}$, $k\ne 0$. 
Here $\alpha_{n,k}$, $k>0$,  is the number of $k$-vertex paths $(\ell_1,\dotsc,\ell_k)$ for which $\ell_1=1$ and $\ell_k=\abs{ 2n}+\theta(n)$. 
Since we have the recurrence    $\alpha_{n,k+1}=\alpha_{n-1,k}+\alpha_{n+1,k}$ and $\alpha_{n,1}=\delta_{n 0}$, it follows via the addition formula for binomial coefficients \cite{Graham1994} that 
\begin{align*}
\alpha_{n,k}=
\begin{cases}
\binom{k-1}{(n+k-1)/2}, & (n+k-1)/2 = 0,1,\dotsc,k-1, \\ 
0, &  \text{otherwise.}  
\end{cases}
\end{align*}
It follows from the definition that $g_n(y,x,t)$ is, up to  parametrization, the discrete propagator for a free random walk process and in particular that   
\begin{gather}\label{eq:freediscreteprop}
g_n(y,x,t)=\alpha_{(nL+y-x)/\sigma,\,t/\uptau+1}2^{-t/\uptau}. 
\end{gather} 
The above relation between $g_n(y,x,t)$ and $q_k(y,x,t)$ implies that, if $n\ge 0$, then  
\begin{align}
&\left[\begin{array}{l}
g_{n}(y,x,t)-g_{n+1}(L-y,x,t)\\
-g_{-n-1}(y,x,t)+g_{-n}(L-y,x,t)
\end{array}\right]\nonumber\\
&\phantom{...............}=\sum_{k\ge 0}^{}
\mat A_{(n,k)}
\left[\begin{array}{l}
q_{k}(y,x,t)\\
q_{k}(y,L-x,t)
\end{array}\right]\label{eq:infiniteequations}, 
\end{align}
where $\mat{A}_{(n,k)}=\delta_{n 0}\mat I$ if $k=0$ and  
\begin{align*}
\mat{A}_{(n,k)}=\left[\begin{array}{cc}
\alpha_{n-1,k}-\alpha_{n+1,k} & \alpha_{n,k}-\alpha_{n+2,k}\\
\alpha_{n,k}-\alpha_{n+2,k} & \alpha_{n-1,k}-\alpha_{n+1,k}
\end{array}\right]
\end{align*}
if $k>0$. Throughout, $\mat I$ is the identity matrix and $\mat 0$ is the zero matrix whose sizes are determined by the context.  
Let $\vec g=\mat\aleph\vec q$, where $\mat\aleph$ is the infinite matrix whose first and second rows are 
$\left[\begin{smallmatrix}\mat{A}_{(0,0)}&\mat {A}_{(0,1)}&\cdots\end{smallmatrix}\right]$, third and fourth rows are  $\left[\begin{smallmatrix}\mat{A}_{(1,0)}&\mat{A}_{(1,1)}&\cdots\end{smallmatrix}\right]$, and so on, be the matrix equation for the system of equations we get by writing \eqref{eq:infiniteequations} with $n=0$, \eqref{eq:infiniteequations} with $n=1$, and so on underneath each other. Since $\mat{A}_{(n,k)}$ is equal to $\mat I$ if $n=k$ and equal to $\mat 0$ if $n>k$,  it follows that $\mat\aleph$ is upper unitriangular. 
The inverse $\mat\aleph^{-1}$ of $\mat\aleph$ is the infinite upper unitriangular matrix whose first and second rows are $\left[\begin{smallmatrix}{\mat B}_{(0,0)}&{\mat B}_{(0,1)}&\cdots\end{smallmatrix}\right]$, third and fourth rows are  $\left[\begin{smallmatrix}{\mat B}_{(1,0)}&{\mat B}_{(1,1)}&\cdots\end{smallmatrix}\right]$, and so on, where 
\begin{align*}
{\mat B}_{(k,n)}=\left[\begin{array}{cc}
\beta_{n+1,k}&-\beta_{n,k}\\-\beta_{n,k}&\beta_{n+1,k}
\end{array}\right]
\end{align*}
and $\beta_{n,k}$, $k\ge 0$, is  
\begin{align*}
\beta_{n,k}=
\begin{cases}
\sum_{m=0}^{(|n|-k-1)/2}\binom{-k}{m}, &\text{} (|n|-k-1)/2 = 0,1,\dotsc,\\ 
0, &  \text{otherwise.}  
\end{cases}
\end{align*}
That this is the inverse, i.e.,     
$\sum_{n\ge 0}^{}{\mat B}_{(k,n)}\mat{A}_{(n,l)}=\delta_{kl}\mat I$, 
follows via the Chu--Vandermonde identity (see the Supplemental Material \cite{Buhlsupplementary}).  
Note that, if $k>0$ and $\beta_{n,k}\ne 0$, then it follows from the identity $\binom{-k}{m}=(-1)^m\binom{k+m-1}{m}$ \cite{Graham1994} that  $\beta_{n,k}$ is an alternating sum of entries in the $k$th diagonal of Pascal's triangle.  
Application of $\mat\aleph^{-1}$ to $\vec g=\mat\aleph\vec q$ gives the  following analytic expression:  
\begin{equation*}
\begin{split} 
q_k(y,x,t)&=\sum_{n}^{}a_{nk}'\,g_n(y,x,t)+ b_{nk}'\,g_n(L -y,x,t), 
\end{split}
\end{equation*}
where $a_{n 0}'=\beta_{n+1,0}$ and $a_{nk}'=\beta_{n+\sgn(k),\abs k}$, $k\ne 0$, together with  $b_{n 0}'=-\beta_{n,0}$ and $b_{nk}'=-\beta_{n,\abs k}$, $k\ne 0$ 
[the calculation uses that     $g_n(y,x,t)=g_{-n}(L-y,L-x,t)$].   

We can now obtain an analytic expression for $K(y,x,t)_{ji}$ as follows:  If $\mathcal G$ is finite ($E,V<\infty$), then we let $\mat\Phi$ and $\mat\Psi$ be the two $E\times V$ matrices defined by         
\begin{align*}
\mat{\Phi}_{ij}=\delta_{{\ori_i} j},\qquad\mat{\Psi}_{i j}=\delta_{{\ter_i} j}.
\end{align*}
The sum of these two matrices is equal to the edge-vertex incidence matrix, i.e., $(\mat\Phi+\mat\Psi)_{ij}$ equals $1$ if $e_i$ has an end point that coincides with $v_j$ and equals $0$ otherwise. In addition, we have that 
\begin{subequations}\label{eq:PhiAD}
\begin{align}
&\mat\Phi^\top\mat{\Phi}+\mat{\Psi}^\top\mat{\Psi}=\mat{D},\label{eq:Phi=D}\\
&\mat\Phi^\top\mat{\Psi}+\mat{\Psi}^\top\mat{\Phi}=\mat{A},\label{eq:Phi=A}
\end{align}
\end{subequations}
where $\mat D$ is the degree matrix, i.e., $\mat{D}_{ij}=d_{i}\delta_{ij}$ and $\mat A$ is the adjacency matrix, i.e., $\mat A_{ij}$ is the number of edges between $v_{i}$ and $v_{j}$. Let 
\begin{align*}
\mat G=\left[\begin{array}{cc}
\mat\Phi\mat C\mat\Psi^\top & \mat\Phi\mat C\mat\Phi^\top\\
\mat\Psi\mat C\mat\Psi^\top & \mat\Psi\mat C\mat\Phi^\top
\end{array}\right],
\qquad 
\mat J=\left[\begin{array}{cc}
\mat 0 & \mat I_{}\\
\mat I_{} & \mat 0
\end{array}\right],
\end{align*}
where $\mat C$ is the $V\times V$ matrix defined by  
$\mat C_{i j}=\frac{2c_{i}}{d_{i}}\delta_{ij}$.   
It follows from \eqref{eq:Phi=A} that the $k$th power of $\mat G$ can be written      
\begin{align*}
\mat G^k=\left[\begin{array}{cc}
\mat\Phi(\mat C\mat A)^{k-1}\mat C\mat\Psi^\top & \mat\Phi(\mat C\mat A)^{k-1}\mat C\mat\Phi^\top\\
\mat\Psi(\mat C\mat A)^{k-1}\mat C\mat\Psi^\top & \mat\Psi(\mat C\mat A)^{k-1}\mat C\mat\Phi^\top
\end{array}\right].
\end{align*}
We have via the definition of matrix multiplication that $\bm{(}\mat\Phi(\mat C\mat A)^{k-1}\mat C\mat\Psi^\top\bm{)}_{ji}\,q_k(y,x,t)$ is equal to that part of the inner sum in \eqref{eq:firstqnexpression} that is over all $k$-vertex paths $(\ell_1,\dotsc,\ell_k)$ for which $\ell_1=\ter_i$ and $\ell_k=\ori_j$. 
The remaining part of the inner sum can be expressed in a similar way using the other three blocks in $\mat G^k$.  Consequently, we have that       
\begin{equation*}
K(y,x,t)_{ji}\!=
\!\!\sum_{k}(\mat G^{\bar k})_{ji}\,q_k(y,x,t) + (\mat J\mat G^{\bar k})_{ji}\,q_k(L-y,x,t),  
\end{equation*}
where $\mat G^{\bar k}$ is equal to $\mat G^k$ if $k\ge 0$ and equal to $(\mat G^\top)^{k}$ if $k<0$.  
By inserting the analytic expression for $q_k(y,x,t)$  and interchanging the order of the summation we get ($\beta_{n,k}=0$ if $|n|\le k$)   
\begin{align*}
K(y,x,t)_{ji}\!=\!\!\sum_{n} 
(\mat S_n)_{ji}\,g_{n}(y,x,t)\!+\!(\mat J\mat S_n)_{ji}\,g_{n}(L\!-\!y,x,t),
\end{align*}
where $\mat S_{n}=\sum_{k}^{}a_{nk}'\mat G^{\bar k}+b_{nk}'\mat J\mat G^{\bar k}$ or, equivalently, 
\begin{subequations}
\begin{align}
\mat S_n&=(-\mat J)^n+\mat H_{n+1}-\mat J\mat H_{n},\label{eq:Sndef}\\
\mat H_{n}&=\sum_{k>0}^{}\beta_{n,k}\mat G^{k}-\beta_{n-1,k}\mat G^{k}\mat J.\label{eq:Hndef}
\end{align}
\end{subequations}
We have via the addition formula for binomial coefficients \cite{Graham1994} that, if $k> 0$, then $\beta_{n,k-1}=\beta_{n-1,k}+\beta_{n+1,k}$. Inserting this in \eqref{eq:Hndef} gives {$\mat H_{n+1}=\mat G(-\mat J)^{n+1}+\mat G\mat H_{n}-\mat H_{n-1}$}, which together with \eqref{eq:Sndef} leads to the following recurrence:
\begin{align}\label{eq:4Ex4Erecir}
\left[\begin{array}{c}
\mat S_{n+1} \\
\mat H_{n+1} 
\end{array}\right]
=\mat T
\left[\begin{array}{c}
\mat S_n \\
\mat H_n 
\end{array}\right]
-\left[\begin{array}{c}
\mat 0 \\
(-\mat J)^n 
\end{array}\right],
\end{align}
where $\mat T$ and the inverse of $\mat T$ are 
\begin{align*}
\mat T=\left[\begin{array}{cc}
\mat G-\mat J & \mat G\mat J -2\mat I \\
\mat I & \mat J
\end{array}\right],
\quad\mat T^{-1}
=\left[\begin{array}{cc}
\mat J & -\mat J\mat G+2\mat I \\
-\mat I & \mat G-\mat J
\end{array}\right].
\end{align*}
By subtracting \eqref{eq:4Ex4Erecir} shifted one down from \eqref{eq:4Ex4Erecir} shifted one up we get $F_{n+2}=\mat T\mat F_{n+1}+\mat F_{n}-\mat T\mat F_{n-1}$, where $\mat F_n=\left[\begin{smallmatrix}\mat S_n\\\mat H_n\end{smallmatrix}\right]$. From this recurrence we get (the $12E\times 2E$ matrix below consists of, from top to bottom,  $\mat F_1$, $\mat F_0$, and $\mat F_{-1}$)
\begin{align}\label{eq:Sn=U^n}
\mat S_n=\left[\begin{array}{cccccc}\mat I & \mat 0 & \mat 0 & \mat 0 & \mat 0 & \mat 0
\end{array}\right]
\begin{matrix}\mat U^{n-1}\end{matrix}
\left[\begin{matrix}\mat G-\mat J\\\mat 0 \\\mat I\\\mat 0\\\mat J\mat G\mat J-\mat J\\-\mat G\mat J
\end{matrix}\right], 
\end{align}
where $\mat U$ and the inverse of $\mat U$ are
\begin{align*}
\mat U=\left[\begin{array}{ccc}
\mat T & \mat I & -\mat T\\
\mat I & \mat 0 & \mat 0\\
\mat 0 & \mat I & \mat 0
\end{array}\right],
\qquad\mat U^{-1}=\left[\begin{array}{ccc}
 \mat 0 & \mat I & \mat 0\\
 \mat 0 & \mat 0 & \mat I\\
-\mat T^{-1} & \mat I & \mat T^{-1}
\end{array}\right].
\end{align*}
If $c_1,\dotsc,c_V=0,1$, then \eqref{eq:Sn=U^n} reduces to $\mat S_n=(\mat G-\mat J)^n$ since in this special case we have that $\mat G\mat J\mat G=2\mat G$ and that $\mat G-\mat J$ is orthogonal (the latter can be seen by using the fact that $\mat G^\top=\mat J\mat G\mat J$).
If $\mathcal G$ is infinite ($E,V=\infty$), then we let $\mat\Phi$, $\mat\Psi$, and $\mat C$ each have a variable (finite) size and we can then use the same procedure as in the finite case provided that a sufficiently large size of each of these matrices is used at each step in the calculation.

If $\uptau\ll t$, then we have via \eqref{eq:freediscreteprop} and the de Moivre--Laplace theorem that $g_n(y,x,t)\simeq 2\sigma\bar g_n(y,x,t)$ at nonzero values of $g_n(y,x,t)$, where $\bar g_n(y,x,t)$ is defined below and $\dcoef=\sigma^2/(2\uptau)$ is the diffusion coefficient. 
Thus, if $\uptau\ll t$ and $\sigma\ll b-a$, where $0\le a<b\le L$, then we have that  $\sum_{a< y< b} K(y,x,t)_{ji}\simeq\int_a^b\bar K(y,x,t)_{ji}\du y$, where 
\begin{gather*}
\bar{K}(y,x,t)_{ji}\!=\!\!\sum_{n}(\mat S_n)_{ji}\,\bar g_{n}(y,x,t)\!+\!(\mat J\mat S_n)_{ji}\,\bar{g}_{n}(L\!-\!y,x,t),\\
\bar{g}_n(y,x,t)=(4\pi\dcoef t)^{-1/2}
\exp\bm{(}-(nL+y-x)^2/(4\dcoef t)\bm{)}
\end{gather*}
is the continuum-limit propagator, i.e., the general probability density function we get by taking the continuum limit of $K(y,x,t)_{ji}$ while keeping the ratio between the step length squared and the step duration fixed (see the Supplemental Material \cite{Buhlsupplementary}) (this way of taking the continuum limit is the same as that used to obtain the diffusion equation and the Wiener or Brownian motion process).            

It follows from the derived expressions that $\bar K(y,x,t)_{ji}$ has the properties described in this paragraph. First,      
\begin{align*}
\lim_{t\rightarrow 0}\bar{K}(y,x,t)_{ji}&=\delta_{ji}\,\delta(y-x),\quad 0<x<L,
\end{align*}
where $\delta(\cdot)$ is the Dirac delta function.   
Let $p$ be a non-negative integer. Using the Hermite polynomial formula for the $p$th derivative of a Gaussian it can be seen that, if $|n|>1$, then  $|\tfrac{\partial^p}{\partial{t}^p}\bar g_n(y,x,t)|\le M_p\dcoef^p  /|nL+y-x|^{2p+1}$, where $M_p$ is a $p$-dependent constant. This bound and the basic theorems on function series can be used to show that $\bar K(y,x,t)_{ji}$ has a partial derivative with respect to $y$, $x$, and $t$ of any order that can be obtained by taking the appropriate differential operator inside the series. Since $\tfrac{\partial}{\partial{t}}\bar g_n(y,x,t)=\dcoef\tfrac{\partial^2}{\partial{y}^2}\bar g_n(y,x,t)$ it consequently follows that $\bar K(y,x,t)_{ji}$ satisfies the diffusion (or heat) equation. 
If $\ori_j=\ell$ or $\ter_j=\ell$, then it can be seen by substituting $n+1$ for $n$ either in $(\mat S_n)_{ji}\,\bar g_{n}(y,x,t)$ or in $(\mat J\mat S_n)_{ji}\,\bar{g}_{n}(L-y,x,t)$ and using \eqref{eq:4Ex4Erecir} that the $p$th normal partial derivative of $\bar K(y,x,t)_{ji}$ with respect to $y$ at  $y=\delta_{\ter_j \ell}L$ can be expressed as     
[the superscript $(p)$ is used to denote the $p$th partial derivative with respect to $y$]
\begin{align}
&(\delta_{\ori_j \ell}-\delta_{\ter_j \ell})^p\bar{K}^{(p)}(\delta_{\ter_j \ell}L,x,t)_{ji}\nonumber\\
&\!=\!\!\sum_{n}
(\bm{(}\mat G\mat S_{n}+(\mat G\mat J-2\mat I)\mat H_n\bm{)}_{m i}\!-\!\gamma_p(\mat S_{n})_{\hat m i})\bar{g}_n^{(p)}(L,x,t),\label{eq:pthnormal}
\end{align}
where $\gamma_p=1-(-1)^p$ and $(m,\hat m)=(j,j+E)$ if $\ori_j=\ell$ and $(m,\hat m)=(j+E,j)$ if $\ter_j=\ell$.  For the $m$ in \eqref{eq:pthnormal} we have that $\mat G_{m k}=\tfrac{2c_\ell}{d_\ell}\mat\Gamma_{\ell k}$ and $(\mat G\mat J\mat G-2\mat G)_{m k}=\tfrac{4c_\ell}{d_\ell}(c_{\ell}-1)\mat\Gamma_{\ell k}$, where $\mat\Gamma=\left[\begin{smallmatrix}\mat\Psi^\top & \mat\Phi^\top\end{smallmatrix}\right]$, which together with \eqref{eq:pthnormal} shows that      
\begin{align*}
&\bar{K}^{(2p)}(\delta_{\ter_j \ell}L,x,t)_{j i}=\bar K^{(2p)}(\delta_{\ter_k \ell}L,x,t)_{k i},\quad j,k\in N_\ell,\\  
&\bar{K}^{(2p)}(\delta_{\ter_j \ell}L,x,t)_{j i}=0,\quad j\in N_\ell,\:\:\: c_\ell=0,\\
&\sum_{j\in N_\ell}(\delta_{\ori_j \ell}-\delta_{\ter_j \ell})\bar{K}^{(2p+1)}(\delta_{\ter_j \ell}L,x,t)_{j i}=0,\quad c_\ell=1,
\end{align*}
where $N_\ell=\{\,j\mid\ori_j=\ell\vee\ter_j=\ell\,\}$. At a vertex whose associated continuation probability is  $0$ (completely absorbing) or $1$ (nonabsorbing) a Dirichlet or a Kirchhoff boundary condition, respectively, is thus satisfied  (the latter reduces to a Neumann boundary condition if the vertex has degree $1$).

If $\mathcal G$ is an infinite $k$-armed star and  $(v_\ell)_{\ell=1}^\infty$, $(e_i)_{i=1}^\infty$, the edge directions, and the continuation probabilities are such that $\ori_1,\dotsc,\ori_k=1$, $\ori_{i+k}=\ter_{i}=i+1$, and  $c_2,c_3,\dotsc=1$, then it can be seen by looking at the structure of $\mat G-\mat J$ that $\bar K(y,x,t)_{ji}$ is equal to  
\begin{align*}
\epsilon\,\bar g_{\jmath-\imath}(y,x,t)+({2c_1}/{k}-\epsilon)\,\bar g_{\jmath+\imath}(y,-x,t),\quad c_1=0,1, 
\end{align*}
where $\imath=\lfloor(i-1)/k \rfloor$, $\jmath=\lfloor(j-1)/k \rfloor$, and $\epsilon$ is equal to $1$ if $(j-i)/k$ is an integer (i.e., if $e_i$ and $e_j$ are part of the same arm) and equal to $0$ otherwise.  This expression entails the well-known method-of-images expansions for the half-space. 
The other known method-of-images expansions follow similarly as special cases.

If $\mathcal G$ is finite and connected, then the formulas derived in the Supplemental Material \cite{Buhlsupplementary} give a positive integer $\mu$, a    $2E\times\mu$ matrix $\mat\Omega$, and a $\mu\times\mu$ diagonal matrix $\mat\Theta$ whose diagonal entries are between $0$ and $2\pi$ such that $\mat S_n=\mat \Omega\ex^{-\iu n\mat\Theta}\mat\Omega^{\dagger}$ and $\mat J\mat\Omega^\ast=\mat\Omega\ex^{-\iu\mat\Theta}$. 
By inserting this above and applying the Poisson summation formula we get, after noting that the entrywise Fourier transform of $\ex^{-\iu n\mat\Theta}\bar{g}_n(y,x,t)$ with respect to $n$ is equal to $L^{-1}\ex^{-(2\pi\xi\mat I+\mat\Theta)^2\dcoef t/L^2 +\iu (2\pi\xi\mat I+\mat\Theta)(y-x)/L}$,  
\begin{gather*}
\bar{K}(y,x,t)_{ji}=\sum_{n\ge 0}^{}\ex^{-k_{n}^2\dcoef t} u_{j n}(y) u_{i n}(x),
\end{gather*}
where $k_n=0$ and $u_{i n}(x)=(EL)^{-1/2}b$ if $n=0$ and  
\begin{gather*}
k_{n}L=\lfloor(n-1)/\mu\rfloor 2\pi +\mat\Theta_{r_n r_n},\\ 
u_{in}(x)=L^{-1/2}\bm{(}({\mat J\mat\Omega^{\ast}})_{i r_n}\ex^{\iu k_{n}x}+(\mat J\mat\Omega)_{i r_n}\ex^{-\iu k_{n}x}\bm{)}
\end{gather*}
if $n>0$ [in the above expressions $b=\lfloor c_1c_2\dotsm c_V\rfloor$ and      
$r_n=n-\lfloor(n-1)/\mu\rfloor\mu$]. 
It follows from the derived expressions that, if $c_1,\dots,c_V=0,1$, then   $\bm{(}u_{in}(x)\bm{)}_{n=1-b}^\infty$ is complete and consists of orthonormal, i.e., $\sum_{i=1}^{E}\int_0^L u_{i m}(x){u_{i n}(x)}\du x=\delta_{m n}$, Laplace eigenfunctions that satisfy the boundary conditions  above.     

\begin{figure}[!htp]
\centering
\includegraphics[]{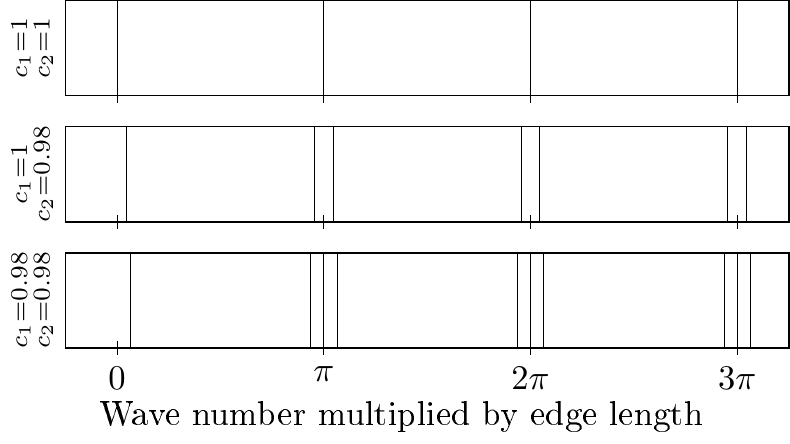}
\caption{\label{fig:linesplitting}The first part of $(k_nL)_{n=1-b}^\infty$ for a geometric graph with only one edge and three sets of values for $c_1$ and $c_2$ [corresponding, e.g., to particle diffusion between two parallel plane objects whose  absorptivities are $0$ and $0$ (top), $0$ and $0.02$ (middle), and $0.02$ and $0.02$ (bottom)].}
\end{figure}

If $\mathcal G$ has only one edge, then we have that $\mu$ is equal to $2$ if $c_1,c_2=0,1$, equal to $6$ if $0<c_1,c_2<1$, and equal to $4$ otherwise and that the wave numbers multiplied by the edge length   $(k_nL)_{n=1-b}^\infty$ are, when arranged in ascending order,    
\begin{align*}
&(0,\pi,2\pi,3\pi,\dotsc),\quad c_1,c_2=1,\\
&(\pi,2\pi,3\pi,4\pi,\dotsc),\quad c_1,c_2=0,\\
&(\tfrac 12\pi,\tfrac 32\pi,\tfrac 52\pi,\tfrac 72\pi,\dotsc),\quad c_1=0,\, c_2=1,\\
&(\tfrac 12\pi,\pi,\tfrac 32\pi,2\pi,\tfrac 52\pi,\dotsc),\quad c_1=0,\, 0<c_2<1,\\  
&(a,\pi-a,{}\pi+a,2\pi-a,2\pi+a,\dotsc),\quad c_1=1,\, 0\!<\!c_2\!<\!1,\\  
&(a,\pi-a,\pi,\pi+a,2\pi-a,2\pi,2\pi+a,\dotsc),\quad  \!0\!<\!c_1,c_2\!<\!1,   
\end{align*}
where $a=\acos(\sqrt{c_1c_2})$ and similarly if $c_1$ and $c_2$ are interchanged (see the Supplemental Material \cite{Buhlsupplementary}). 
The first three of these sequences are known from the traditional approach, i.e., in this specific case, from  the solving of Sturm--Liouville problems. The last three sequences, on the other hand, cannot occur with the traditional approach and show, in particular, how the presence of partially absorbing barriers can lead to line splitting in the wave-number spectrum (see, e.g., Fig.~\ref{fig:linesplitting}).   

\begin{figure}[!t]
\vspace{0.38cm}
\centering
\includegraphics[]{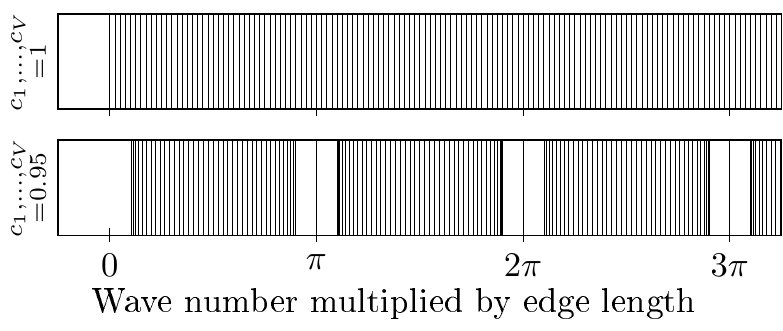}
\setlength{\belowcaptionskip}{-0.44cm}\caption{\label{fig:bandgap}The first part of $(k_nL)_{n=1-b}^\infty$ for a chain with 40 edges (i.e., a geometric graph with two degree-$1$ and 39 degree-$2$   vertices) and two sets of values for  $c_1,\dotsc,c_V$ [corresponding, e.g., to particle diffusion in an array of 41 parallel plane barriers where each of the 39 inner barriers reflect and transmit equally and all the barriers have an absorptivity of $0$ (top) and $0.05$ (bottom)].    
}
\end{figure}

If $\mathcal G$ is finite and connected as well as bipartite and $c_1,\dotsc,c_V=\varkappa$, where $0<\varkappa <1$, then it follows that the sorted unique values in the main diagonal of $\mat\Theta$ form a sequence of the form    $(a,\dotsc,\pi-a,\pi,\pi+a,\dotsc,2\pi-a,2\pi)$, where $a=\acos(\varkappa)$ and thereby  that $(k_n)_{n=1-b}^\infty$ will have band gaps or forbidden bands (see, e.g., Fig.~\ref{fig:bandgap}).

To summarize, we have shown that the (free random walk) central limit theorem can be generalized to cover also the nonfree case. We have hereby obtained a continuum-limit description of the diffusive motion where the effect of partially absorbing barriers is accounted for in a natural and non-Markovian way that, in contrast to the traditional approach, quantifies the absorptivity of a barrier in terms of a dimensionless parameter in the range $0$ to $1$.  
Given that the continuum-limit description rests directly on the normal approximation, it seems reasonable to expect that it will be able to describe instances of the considered physical phenomenon more quantitatively accurately than the traditional continuum-limit description.
Additional aspects to investigate or consider in connection with the approach include the following: The factorization of the continuum-limit propagator for rectangular cuboid domains (using the Kronecker product) corresponding to a separation of spatial variables.
The determination of the form of the continuum-limit propagator in the case where, e.g., for a chain with $m$ edges each of length $L/m$ and a continuation probability of $1$ for all the degree-$2$ vertices, $m$ is made to go to infinity and the continuation probabilities for the two degree-$1$ vertices to $1$ as  part of taking the continuum limit.  
It would also be interesting to try to generalize the approach to allow for systematically biased random displacements and a nonzero angular frequency in each of the plane waves---what could the wave-number spectrum  resemble in this case?

This work was partially funded by the FP7 EU grant-funded  project AirPROM.


%





\pagebreak
\widetext
\begin{center}
\textbf{\large Different approach to the modeling of nonfree particle diffusion:\\ Supplemental Material}
\author{Niels Buhl}
\end{center}

\setcounter{equation}{0}
\setcounter{figure}{0}
\setcounter{table}{0}
\setcounter{page}{1}
\makeatletter
\renewcommand{\theequation}{S\arabic{equation}}
\renewcommand{\thefigure}{S\arabic{figure}}
\renewcommand{\bibnumfmt}[1]{[S#1]}
\renewcommand{\citenumfont}[1]{S#1}

\vspace{-0.2cm}
\section{Section}
In this section we prove that $\big(\sum_{n\ge 0}^{}{\mat B}_{(k,n)}\mat{A}_{(n,l)}\big)_{ij}=\delta_{kl}\delta_{ij}$. 
Assume first that $i\ne j$. If $l=0$, then we have that $\big(\sum_{n\ge 0}^{}{\mat B}_{(k,n)}\mat A_{(n,l)}\big)_{ij}=0$ 
and if $l>0$, then we have that  
\begin{align*}
\big(\sum_{n\ge 0}^{}{\mat B}_{(k,n)}\mat{A}_{(n,l)}\big)_{ij}
&=\sum_{n\ge 0}^{}-\beta_{n,k}(\alpha_{n-1,l}-\alpha_{n+1,l})+\beta_{n+1,k}(\alpha_{n,l}-\alpha_{n+2,l})
\\&
=-\beta_{0,k}(\alpha_{-1,l}-\alpha_{1,l})=0.
\end{align*}
Assume next that $i=j$. If $l=0$, then we have that $\big(\sum_{n\ge 0}^{}{\mat B}_{(k,n)}\mat A_{(n,l)}\big)_{ij}=\delta_{k 0}$      
and if $k=2p$ and $l=2q$, where $p$ is a non-negative integer and $q$ is a positive integer, then we have that     
\begin{align*}
\big(\sum_{n\ge 0}^{}{\mat B}_{(k,n)}\mat{A}_{(n,l)}\big)_{ij}
=&\sum_{n\ge 0}^{}\beta_{n+1,2p}(\alpha_{n-1,2q}-\alpha_{n+1,2q})-\beta_{n,2p}(\alpha_{n,2q}-\alpha_{n+2,2q})\\
=&\sum_{n\ge 0}^{}(\beta_{2n+1,2p}-\beta_{2n-1,2p})(\alpha_{2n-1,2q}-\alpha_{2n+1,2q})\\
=&\sum_{n\ge 0}^{}\binom{-2p}{-p+n}\left(\binom{2q-1}{q-n}-\binom{2q-1}{q-1-n}\right)\\
=&\binom{2q-2p-1}{q-p}-\binom{2q-2p-1}{q-p-1}=\delta_{pq},
\end{align*}
where we have used  $\alpha_{-1,2q}-\alpha_{1,2q}=0$, the symmetry identity \cite{Graham1994sup}, and the Chu--Vandermonde identity \cite{Graham1994sup}
\begin{align*}
\sum_{k}\binom{r}{m+k}\binom{s}{n-k}=\sum_{k}\binom{r}{m+k}\binom{s}{m+n-(m+k)}=\binom{r+s}{m+n}.
\end{align*} 
If $k=2p+1$ and $l=2q+1$, where $p$ and $q$ are non-negative integers, then we have that 
\begin{align*}
\big(\sum_{n\ge 0}^{}{\mat B}_{(k,n)}\mat{A}_{(n,l)}\big)_{ij}
=&\sum_{n\ge 0}^{}\beta_{n+1,2p+1}(\alpha_{n-1,2q+1}-\alpha_{n+1,2q+1})-\beta_{n,2p+1}(\alpha_{n,2q+1}-\alpha_{n+2,2q+1})\\
=&\sum_{n\ge 0}^{}(\beta_{2n+2,2p+1}-\beta_{2n,2p+1})(\alpha_{2n,2q+1}-\alpha_{2n+2,2q+1})\\
=&\sum_{n\ge 0}^{}\binom{-2p-1}{-p+n}\left(\binom{2q}{q+n}-\binom{2q}{q+n+1}\right)\\
=&\binom{2q-2p-1}{q-p}-\binom{2q-2p-1}{q-p-1}
=\delta_{pq}.
\end{align*}
If $k$ and $l$ have different parity and $l>0$, then we have that $\big(\sum_{n\ge 0}^{}{\mat B}_{(k,n)}\mat A_{(n,l)}\big)_{ij}=0$, which completes the proof.

\vspace{-0.2cm}
\section{Section}
In this section we show formally that $\bar K(y,x,t)_{ji}$ is the general probability density function we get by taking the continuum limit of $K(y,x,t)_{ji}$ while keeping the ratio between the step length squared and the step duration fixed.  
Let $(\xi_l)_{l=1}^\infty$ be statistically independent random variables such that $P(\xi_l=-1)=P(\xi_l=1)=1/2$.  
Substitute $\uptau/m^2$ for $\uptau$ and $\sigma/m$ for $\sigma$ throughout and restrict the values that $x$ and $t$ can take to those that were possible before the two substitutions. 
The definition of $g_n(y,x,t)$ then gives the first equality below and the (free random walk) central limit theorem the second  
\begin{gather*}
\lim_{m\rightarrow\infty}\sum_{y<b} g_n(y,x,t)=\lim_{m\rightarrow\infty}\sum_{y<b}P(\tfrac{\sqrt{t/\uptau}}{\sqrt{m^2t/\uptau}}\sum_{l=1}^{m^2t/\uptau}\sigma\xi_l=nL+y-x)
=\int_{-\infty}^b\bar g_n(y,x,t)\du y
\end{gather*}
since the variance of $\sqrt{\sigma^2 t/\uptau}\xi_l$ is equal to $\sigma^2t/\uptau=2\dcoef t$. With the help of this result we finally get  
\begin{gather*}
\lim_{m\rightarrow\infty}\sum_{a<y< b} K(y,x,t)_{ji}=\int_a^b\bar K(y,x,t)_{ji}\du y.
\end{gather*}

\section{Section}
In this section we show in full detail how to, if $\mathcal G$ is finite and connected, get the 'infinite sum of plane waves' expression for $\bar K(y,x,t)_{ji}$ via formulas that are  directly implementable with standard mathematical software.
Assume thus that $\mathcal G$ is finite and connected.  
Let $\mat W$ be the $\rho\times\rho$ matrix    
\begin{align*}
\mat W={\mat R}^\top\mat A\mat R,
\end{align*}
where $\mat R$ is the matrix we get by deleting all columns that consist entirely of zeros from $(\tfra 2\mat C)^{1/2}$  and $\rho$ is the number of columns in $\mat R$.
It follows that $\mat C=2\mat R\mat R^\top$.  
Let ${\mat V}$ and ${\mat X}$ be real $\rho\times\rho$ matrices such that $\mat W{\mat V}={\mat V}{\mat X}$, ${\mat V}^\top{\mat V}=\mat I$, and ${\mat X}$ is diagonal with the diagonal entries arranged in ascending order (we can find these matrices since $\mat W$ is real symmetric and therefore real orthogonally diagonalizable).     
Let $\vec a$ be the column vector corresponding to the $j$th column in $\mat R{\mat V}$. Since $\mat A=-\mat D+(\mat\Phi+\mat\Psi)^\top(\mat\Phi+\mat\Psi)$ and $\mat A=\mat D-(\mat\Phi-\mat\Psi)^\top(\mat\Phi-\mat\Psi)$ we have that  
\begin{align}\label{eq:eigenvaluebounds}
-1\le -\vec a^\top\mat D\vec a+\sum_{i=1}^{E}({\vec a_{\ori_i}} +\vec a_{\ter_i})^2={\mat X}_{j j}=\vec a^\top\mat D\vec a-\sum_{i=1}^{E}(\vec a_{\ori_i}-\vec a_{\ter_i})^2\le 1.
\end{align} 
It follows from \eqref{eq:eigenvaluebounds} and the assumed connectedness that $1$ occurs in $\mat X$ one time if the condition $c_1,\dotsc,c_V=1$ is satisfied and zero times otherwise. Similarly, it follows that $-1$ occurs in $\mat X$ one time if the conditions $c_1,\dotsc,c_V=1$ and $\mathcal G$ is bipartite are both satisfied and zero times otherwise. 
Let $\hat{\mat V}_-$, $\hat{\mat V}$, $\hat{\mat V}_+$, $\hat{\mat X}_-$, $\hat{\mat X}$, and $\hat{\mat X}_+$ be defined by Table \ref{tab:index} and the two requirements ${\mat V}=\left[\begin{smallmatrix}\hat{\mat V}_- & \hat{\mat V}\, & \hat{\mat V}_+\end{smallmatrix}\right]$ and ${\mat X}=\diag(\hat{\mat X}_-,\hat{\mat X},\hat{\mat X}_+)$. From the above we have that $\mat W\hat{\mat V}=\hat{\mat V}\hat{\mat X}$ and that all entries in $\hat{\mat X}$ are strictly between $-1$ and $1$. 
Let then (the asterisk symbol is used to denote the complex conjugate and the dagger symbol the complex conjugate transpose)  
\begin{align*}
\mat P=\left[\begin{array}{cc}
\mat\Phi\iu\mat R\hat{\mat V}({\sqrt2}\mat Y)^{-1} \\
\mat\Psi\iu\mat R\hat{\mat V}({\sqrt2}\mat Y)^{-1}    
\end{array}\right], 
\qquad\mat Y=(\mat I-\hat{\mat X}^2)^{1/2},     
\end{align*}
\begin{align*}
\begin{aligned}
\mat\Xi&=\left[\begin{array}{cccc}
\mat P-\mat J\mat P\mat Z & \mat P-\mat J\mat P\mat Z^\ast  
\end{array}\right],
\\\mat\Upsilon&=\left[\begin{array}{ccccc} 
\symcen{\mat P-\mat J\mat P\mat Z}{\mat P\mat Z} & \symcen{\mat P-\mat J\mat P\mat Z^\ast}{\mat P\mat Z^\ast} \end{array}\right],
\end{aligned}
\qquad\mat Z=\hat{\mat X}+\iu\mat Y,
\end{align*}
\begin{align*}
\begin{aligned}
\mat M_m=&-\tfra2(\mat\Xi\mat\Xi^\dagger+ m(\mat\Xi\mat\Delta^\ast\mat\Xi^\dagger -\mat J\mat G\mat J+\mat J)-\mat I),
\\\mat N_m=&-\tfra2(\mat\Upsilon\mat\Xi^\dagger+ m(\mat\Upsilon\mat\Delta^\ast\mat\Xi^\dagger +\mat G\mat J )),  
\end{aligned}
\qquad\mat\Delta=\diag(\mat Z^\ast,\mat Z),
\end{align*}
\begin{align*}
\mat Q_m=
\begin{cases}
\bigg[\begin{array}{cc}
\mat\Phi\mat R\hat{\mat V}_m & \phantom{-m}\iu\mat O_m  \\
\mat\Psi\mat R\hat{\mat V}_m & -m\iu\mat O_m   
\end{array}\bigg], &  \text{} c_1,\dotsc,c_V=0,1,\\ 
\iu\mat\daleth_m\mat\gimel_m,  &  \text{otherwise,} 
\end{cases}
\end{align*} 
where $m=-,+$ and $\mat O_m$ is a real matrix of size $E\times \nu_m$ such that $\mat R^\top(\mat\Psi-m\mat\Phi)^\top\mat O_m={\mat 0}$, $\mat O_m^\top\mat O_m=\tfra2\mat I$, and $\nu_m$ is the nullity of $\mat R^\top(\mat\Psi-m\mat\Phi)^\top$. The matrix $\mat\daleth_m$ is a real $2E\times 2E$ matrix such that $\mat M_m\mat\daleth_m=\mat\daleth_m\mat\beth_m$ and  $\mat\daleth_m^\top\mat\daleth_m=\mat I$, where $\mat\beth_m$ is a real $2E\times 2E$ diagonal matrix (we can find  these matrices since $\mat M_m$ is real symmetric as shown below). The matrix $\mat\gimel_m$ is the matrix we get by deleting all columns that consist entirely of zeros from $(\mat\beth_m)^{1/2}$. Let finally   
\begin{align*}
\mat\Omega&=\left[\begin{array}{cccccc}
\mat P-\mat J\mat P\mat Z & \mat Q_- & \mat P-\mat J\mat P\mat Z^\ast & \mat Q_{+} 
\end{array}\right],\\
\mat\Theta&=\diag({\mat\Lambda}, \pi\mat I_{-}, 2\pi\mat I-\mat\Lambda, 2\pi\mat I_{+}), 
\end{align*}
and $\mu$ be the number of columns in $\mat\Omega$, where $\mat\Lambda_{i j}=\acos(\hat{\mat X}_{i i})\delta_{i j}$ 
and $\mat I_m$ is the identity matrix with the same number of columns as $\mat Q_m$. We then have that      
\begin{align}\label{eq:Sn=OmegaexpOmega}
\mat S_n=\mat\Omega\ex^{-\iu n\mat\Theta}\mat\Omega^{\dagger},\qquad \mat J\mat\Omega^\ast=\mat\Omega\ex^{-\iu\mat\Theta}.
\end{align}
That \eqref{eq:Sn=OmegaexpOmega} holds follows from the fact that it  holds in each of the two cases below ($\ex^{-\iu\mat\Theta}=\diag(\mat Z^\ast,-\mat I_-,\mat Z,\mat I_+)$).  
\\Case 1: $c_1,\dotsc,c_V=0,1$. We have in this case that $\mat R^\top\mat D\mat R=\mat I$ and thereby that
\begin{align*}
(\mat G-\mat J)(\mat P-\mat J\mat P\mat Z)&=2\mat P\hat{\mat X}-\mat J\mat P -2\mat P\mat Z +\mat P\mat Z=(\mat P-\mat J\mat P\mat Z)\mat Z^\ast,\\
(\mat G-\mat J)\mat Q_m &=m\mat Q_m, 
\end{align*}
which shows that $(\mat G-\mat J)\mat\Omega=\mat\Omega\ex^{-\iu\mat\Theta}$. 
The rank nullity theorem and the fact that the rank of a matrix is invariant under transposition gives  $\nu_m=E-\rho+\nu_m'$, where $\nu_m'$ is the nullity of $(\mat\Psi-m\mat\Phi)\mat R$.   Since $\nu_m'$ is equal to the number of columns in $\hat{\mat V}_m$ it follows that $\mu=2E$.
We further have that $\mat\Omega$ is unitary $\mat\Omega^\dagger\mat\Omega=\mat\Omega\mat\Omega^\dagger=\mat I$ since
\begin{align*}
(\mat P-\mat J\mat P\mat Z)^{\dagger}(\mat P-\mat J\mat P\mat Z)
&=\tfra2\mat Y^{-2}(\mat I -\hat{\mat X}\mat Z^{\ast} -\hat{\mat X}\mat Z+\mat I)=\mat I, \\
(\mat P-\mat J\mat P\mat Z^\ast)^{\dagger}(\mat P-\mat J\mat P\mat Z)
&=\tfra2\mat Y^{-2}(\mat I -\hat{\mat X}\mat Z-\hat{\mat X}\mat Z +\mat Z^2)=\mat 0,\\ 
(\mat P-\mat J\mat P\mat Z)^\dagger\mat Q_m&=\mat 0,\\
{\mat Q_{n}}^{\dagger}\mat Q_{m}&=\delta_{n m}\mat I,
\end{align*}
which together with the above shows that \eqref{eq:Sn=OmegaexpOmega} holds in this case.    
\\Case 2: $\neg(c_1,\dotsc,c_V=0,1)$. We have in this case that $\mat R\hat{\mat V}\hat{\mat V}^\top\mat R^\top=\tfra2\mat C$ and thereby that  
\begin{subequations}\label{eq:diff}
\begin{align}
\mat\Xi\mat\Delta\mat\Xi^\dagger-\mat\Xi\mat\Delta^\ast\mat\Xi^\dagger&=2\Re((\mat P-\mat J\mat P\mat Z)(\mat Z^\ast-\mat Z)(\mat P-\mat J\mat P\mat Z)^{\dagger}) 
=\mat G -\mat J\mat G\mat J,\\   
\mat\Upsilon\mat\Delta\mat\Xi^\dagger-\mat\Upsilon\mat\Delta^\ast\mat\Xi^\dagger&=2\Re(\mat P\mat Z(\mat Z^\ast-\mat Z)(\mat P-\mat J\mat P\mat Z)^\dagger)=\mat G\mat J,
\\\mat J\mat\Xi\Xi^\dagger+\Xi\mat\Delta^\ast\mat\Xi^\dagger
&=2\Re((\mat J\mat P-\mat J\mat P\mat Z^2)(\mat P-\mat J\mat P\mat Z)^\dagger)=\mat J\mat G\mat J,
\end{align}
\end{subequations}
where $\Re(\cdot)$ is the real part function.
It follows from  \eqref{eq:diff} that $\mat M_m$ is real symmetric, has only non-negative eigenvalues, and satisfies $\mat J\mat M_m=-m\mat M_m$, which implies that $\mat J\mat Q_m^\ast=m\mat Q_m$. We further have that 
\begin{align*}
\mat T\left[\begin{array}{ccccc}
\mat\Xi \\
\mat\Upsilon  
\end{array}\right]
=\left[\begin{array}{cc}
\mat G-\mat J & \mat G\mat J -2\mat I \\
\mat I & \mat J
\end{array}\right]
\left[\begin{array}{ccccc}
\mat P-\mat J\mat P\mat Z & \mat P-\mat J\mat P\mat Z^\ast\\
\mat P\mat Z & \mat P\mat Z^\ast 
\end{array}\right]
=\left[\begin{array}{ccccc}
\mat\Xi \\
\mat\Upsilon  
\end{array}\right]
\mat\Delta, 
\end{align*}
which together with \eqref{eq:diff} gives  
\begin{align*}
\mat T\left[\begin{array}{ccccc}
\mat M_- \\
\mat N_-  
\end{array}\right]
=\left[\begin{array}{cc}
\mat G-\mat J & \mat G\mat J -2\mat I \\
\mat I & \mat J
\end{array}\right]
\left[\begin{array}{ccccc}
-\tfra2(\mat\Xi\mat\Xi^\dagger-\mat\Xi\mat\Delta^\ast\mat\Xi^\dagger +\mat J\mat G\mat J-\mat J-\mat I)\\
-\tfra2(\mat\Upsilon\mat\Xi^\dagger-\mat\Upsilon\mat\Delta^\ast\mat\Xi^\dagger -\mat G\mat J)\phantom{xxxxxxx}
\end{array}\right]
=-\left[\begin{array}{ccccc}
\mat M_-\\
\mat N_-  
\end{array}\right]
+\left[\begin{array}{ccccc}
\mat 0 \\
\tfra2(\mat I+\mat J)  
\end{array}\right],\\
\mat T\left[\begin{array}{ccccc}
\mat M_+ \\
\mat N_+  
\end{array}\right]
=\left[\begin{array}{cc}
\mat G-\mat J & \mat G\mat J -2\mat I \\
\mat I & \mat J
\end{array}\right]
\left[\begin{array}{ccccc}
-\tfra2(\mat\Xi\mat\Xi^\dagger+\mat\Xi\mat\Delta^\ast\mat\Xi^\dagger -\mat J\mat G\mat J+\mat J-\mat I)\\
-\tfra2(\mat\Upsilon\mat\Xi^\dagger+\mat\Upsilon\mat\Delta^\ast\mat\Xi^\dagger +\mat G\mat J)\phantom{xxxxxxx}
\end{array}\right]
=\phantom{-}\left[\begin{array}{ccccc}
\mat M_+ \\
\mat N_+  
\end{array}\right]
+\left[\begin{array}{ccccc}
\mat 0 \\
\tfra2(\mat I-\mat J)  
\end{array}\right].
\end{align*}
By combining the three equations above we get the recurrence 
\begin{align*}
\mat T\left[\begin{array}{ccccc}
\mat\Xi\mat\Delta^n\mat\Xi^\dagger +(-1)^n\mat M_- +\mat M_+\\
\mat\Upsilon\mat\Delta^n\mat\Xi^\dagger +(-1)^n\mat N_- +\symcen{\mat N_+}{\mat N_+}
\end{array}\right]
-\left[\begin{array}{ccc}
\mat 0\\
(-\mat J)^n
\end{array}\right]
=\left[\begin{array}{ccc}
\mat\Xi\mat\Delta^{n+1}\mat\Xi^\dagger +(-1)^{n+1}\mat M_- +\mat M_+\\
\mat\Upsilon\mat\Delta^{n+1}\mat\Xi^\dagger +(-1)^{n+1}\mat N_- +\symcen{\mat M_+}{\mat N_+}
\end{array}\right],
\end{align*}
which together with $\mat\Omega\ex^{-\iu n\mat\Theta}\mat\Omega^{\dagger}=\mat\Xi\mat\Delta^n\mat\Xi^\dagger +(-1)^n\mat M_- +\mat M_+$ and the fact that  $\mat\Xi\mat\Delta^0\mat\Xi^\dagger +(-1)^0\mat M_- +\mat M_+=\mat I$ and $\mat\Upsilon\mat\Delta^0\mat\Xi^\dagger+(-1)^0\mat N_- +\mat N_+=\mat 0$ shows that \eqref{eq:Sn=OmegaexpOmega} holds in this case.

From the 'infinite sum of Gaussians' expression for $\bar K(y,x,t)_{ji}$,  
the fact that $\mat S_{-n}=\mat J\mat S_n\mat J$, and \eqref{eq:Sn=OmegaexpOmega} we get the first equality below. By applying the Poisson summation formula we get the second equality, after noting that the entrywise Fourier transform ($\ex^{-2\pi\iu\xi\cdot}$ kernel) of $\ex^{-\iu n\mat\Theta}\bar{g}_n(y,x,t)$ with respect to $n$ is equal to 
$L^{-1}\ex^{-\sbTheta_\xi^2\dcoef t/L^2 +\iu\sbTheta_\xi(y-x)/L}$,  
where $\sbTheta_\xi=2\pi\xi\mat I+\mat\Theta$, 
\begin{align*}
\bar{K}(y,x,t)_{ji}&=\sum_{n}(\mat J\mat\Omega\ex^{-\iu n\mat\Theta}\mat\Omega^{\dagger}\mat J\bar g_{n}(-y,-x,t) +{\mat J\mat\Omega}\ex^{-\iu n\mat\Theta }\mat\Omega^{\dagger}\bar g_n(L-y,x,t))_{ji}\nonumber\\
&=\sum_{n}\sum_{r=1}^\mu\tfra L(\ex^{-\sbTheta_n^2\dcoef t/L^2})_{rr}(\mat J\mat\Omega\ex^{-\iu\sbTheta_ny/L})_{jr}(\mat J\mat\Omega^\ast\ex^{\iu{\sbTheta}_nx/L}+\mat J\mat\Omega\ex^{-\iu\sbTheta_nx/L})_{ir}\\
&=\sum_{n\ge 0}^{}\ex^{-k_{n}^2\dcoef t} u_{j n}(y) u_{i n}(x),	
\end{align*}
where $k_n=0$ and $u_{i n}(x)=(EL)^{-1/2}\lfloor  c_1c_2\dotsm c_V\rfloor$ if $n=0$ and  
\begin{gather*}
k_{n}L=\lfloor(n-1)/\mu\rfloor 2\pi +\mat\Theta_{r_n r_n},\\ 
u_{in}(x)=L^{-1/2}\bm{(}(\mat J\mat\Omega^{\ast})_{i r_n}\ex^{\iu k_{n}x}+ ({\mat J\mat\Omega^{}})_{i r_n}\ex^{-\iu k_{n}x}\bm{)}
\end{gather*}
if $n>0$ [in the above   expressions   $r_n=n-\lfloor(n-1)/\mu\rfloor\mu$]. To see the third equality, write the blocks in $\dotsb\begin{smallmatrix}{\sbTheta_{-1}}\\\mat\Omega\end{smallmatrix}\begin{smallmatrix}{\sbTheta_0}\\\mat\Omega\end{smallmatrix}\begin{smallmatrix}{\sbTheta_1}\\\mat\Omega\end{smallmatrix}\dotsb$ as shown below and note the symmetry around $\begin{smallmatrix}{0\mat I_+}\\\mat Q_+\end{smallmatrix}$ and that the sum of the terms corresponding to $\begin{smallmatrix}{0\mat I_+}\\\mat Q_+\end{smallmatrix}$  is  equal to  $(EL)^{-1}\lfloor  c_1c_2\dotsb c_V\rfloor$. 
The latter follows from the fact that $\mat Q_+^\ast+\mat Q_+$ has nonzero entries if and only if $c_1,\dotsc,c_V=1$ 
and the fact that $\hat{\mat V}_+=\pm\tfrac{ 1}{\sqrt{2E}}\left[\begin{smallmatrix}\sqrt{d_1}& \cdots&\sqrt{d_V}\end{smallmatrix}\right]^\top$ 
if $c_1,\dotsc,c_V=1$ 
(the factor $\tfra{\sqrt{2E}}$ follows via  the degree-sum formula $\sum_{\ell=1}^V d_\ell=2E$).    
\begin{align*}
\begin{array}{c}\dotsb\! \end{array}
\begin{array}{cccccccccccccc} 
 -{(2\pi\mat I+\mat\Lambda)} & -{2\pi\mat I_+} & -{(2\pi\mat I-\mat\Lambda)} & -{\pi\mat I_-} & -{\mat\Lambda} & {0\mat I_+} & {\mat\Lambda} & {\pi\mat I_-} &  {2\pi\mat I-\mat\Lambda} &  {2\pi\mat I_+} & {2\pi\mat I+\mat\Lambda} \\
 \mat P-\mat J\mat P\mat Z^\ast & \mat Q_+ & \mat P-\mat J\mat P\mat Z & \mat Q_{-} & \mat P-\mat J\mat P\mat Z^\ast & \mat Q_+ & \mat P-\mat J\mat P\mat Z & \mat Q_{-} & \mat P-\mat J\mat P\mat Z^\ast & \mat Q_{+} & \mat P-\mat J\mat P\mat Z 
\end{array} 
\begin{array}{c}\!\!\dotsb \end{array}
\end{align*}

\section{Section}
In this section we show that,  if $\mathcal G$ has only one edge, then the above formula for $\mat\Theta$ gives the following:  
\begin{itemize}
\item If $c_1,c_2=1$, then we have  that $\hat{\mat V}=\left[\begin{smallmatrix}\end{smallmatrix}\right]$, $\hat{\mat X}=\left[\begin{smallmatrix}\end{smallmatrix}\right]$, $\mat Q_-=\tfrac{\pm 1}{\sqrt2}\left[\begin{smallmatrix}{}
-1 \\ 1 \end{smallmatrix}\right]$, and $\mat Q_+=\tfrac{\pm 1}{\sqrt2}\left[\begin{smallmatrix}{}
1  \\ 1 \end{smallmatrix}\right]$. 
This gives $\mat\Theta=\diag(\pi,2\pi)$. 

\item If $c_1,c_2=0$, then we have  that $\hat{\mat V}=\left[\begin{smallmatrix}\end{smallmatrix}\right]$, $\hat{\mat X}=\left[\begin{smallmatrix}\end{smallmatrix}\right]$,
 $\mat Q_-=\tfrac{\pm\iu}{\sqrt2}\left[\begin{smallmatrix}{}
1 \\ 1 \end{smallmatrix}\right]$, and $\mat Q_+=\tfrac{\pm\iu}{\sqrt2}\left[\begin{smallmatrix}{}
1 \\-1 \end{smallmatrix}\right]$. 
This gives $\mat\Theta=\diag(\pi,2\pi)$.

\item If $c_1=0$ and $c_2=1$ or vice versa, then we have that $\hat{\mat V}=\left[\begin{smallmatrix}\pm 1\end{smallmatrix}\right]$, $\hat{\mat X}=\left[\begin{smallmatrix}0\end{smallmatrix}\right]$, $\mat Q_-=\left[\begin{smallmatrix}\end{smallmatrix}\right]$, and $\mat Q_+=\left[\begin{smallmatrix}\end{smallmatrix}\right]$. This gives $\mat\Theta=\diag(\tfrac12\pi,\tfrac32\pi)$.  

\item If $c_1=0$ and $0<c_2<1$ or vice versa, then we have  that $\hat{\mat V}=\left[\begin{smallmatrix}\pm 1\end{smallmatrix}\right]$ and $\hat{\mat X}=\left[\begin{smallmatrix}0\end{smallmatrix}\right]$, which gives $\mat\Xi\mat\Xi^\dagger-\mat I=-2s\mat I$ and   $\mat M_m=s(\mat I-m\mat J)$, where $s=(1-c_1-c_2)/2$. 
The eigenvalues of $\mat M_m$ are $0$ and $2s$. This gives $\mat\Theta=\diag(\tfrac12\pi,\pi,\tfrac32\pi,2\pi)$. 

\item If $0<c_1,c_2<1$, then we have that  
 \begin{alignat*}{3}
\mat W=\left[\begin{array}{cc}
0 & \sqrt{c_1c_2}\\ \sqrt{c_1c_2} & 0
\end{array}\right],\qquad \hat{\mat V}=\left[\begin{array}{cc}
-f & h\\f & h
\end{array}\right],\qquad 
&\hat{\mat X}=\left[\begin{array}{cc}
-\sqrt{c_1c_2} & 0\\0 & \sqrt{c_1c_2}
\end{array}\right],\qquad f,h=\pm\tfrac{1}{\sqrt 2},
\end{alignat*}
which gives $\mat\Xi\mat\Xi^\dagger-\mat I=-2s\mat I$ and $\mat M_m=s(\mat I-m\mat J)$, where $s=(1-c_1)(1-c_2)/(2-2c_1c_2)$. The eigenvalues of $\mat M_m$ are $0$ and $2s$.  
This gives  $\mat\Theta=\diag(\pi-a,a,\pi,\pi+a,2\pi-a,2\pi)$, where $a=\acos(\sqrt {c_1c_2})$.

\item If $c_1=1$ and $0<c_2<1$ or vice versa, then $\hat{\mat V}$ and $\hat{\mat X}$ are as in the previous case and $\mat M_m=\mat 0$. 
This gives  $\mat\Theta=\diag(\pi-a,a,\pi+a,2\pi-a)$, where $a=\acos(\sqrt {c_1c_2})$. 
\end{itemize}

\begin{table}[b]
\begin{ruledtabular}
\begin{tabular}{lcccc}
  & $\hat{\mat V}_-$ & $\hat{\mat V}_+$ & $\hat{\mat X}_-$ & $\hat{\mat X}_+$ \\\hline
$c_1,\dotsc,c_V=1$ and $\mathcal G$ is bipartite  & First column in ${\mat V}$ & Last column in ${\mat V}$ & $\left[\begin{smallmatrix}-1\end{smallmatrix}\right]$ & $\left[\begin{smallmatrix}1\end{smallmatrix}\right]$ \\ 
$c_1,\dotsc,c_V=1$ and $\mathcal G$ is nonbipartite & $\left[\begin{smallmatrix}\end{smallmatrix}\right]$ & Last column in ${\mat V}$ & $\left[\begin{smallmatrix}\end{smallmatrix}\right]$ & $\left[\begin{smallmatrix}1\end{smallmatrix}\right]$ \\ 
$\neg(c_1,\dotsc,c_V=1)$   & $\left[\begin{smallmatrix}\end{smallmatrix}\right]$ & $\left[\begin{smallmatrix}\end{smallmatrix}\right]$ & $\left[\begin{smallmatrix}\end{smallmatrix}\right]$ & $\left[\begin{smallmatrix}\end{smallmatrix}\right]$ 
\end{tabular}
\end{ruledtabular}
\caption{\label{tab:index}
$\left[\begin{smallmatrix}\end{smallmatrix}\right]$ is the $n\times 0$ empty matrix, where $n$ is determined by the context. A geometric graph is bipartite if it is possible to assign either $-1$ or $1$ to each vertex in such a way that each edge is between a vertex assigned $-1$ and a vertex assigned $1$. 
}
\end{table}

%

\end{document}